\documentclass[showpacs,aps,prl,twocolumn]{revtex4}
\usepackage{graphicx}
\usepackage{amsmath}
\usepackage{amssymb}

\begin{document}
\bibliographystyle{apsrev}

\title{Modification of the ground state in Sm-Sr manganites by oxygen
isotope substitution}
\author{N.~A.~Babushkina and E.~A.~Chistotina}
\affiliation{Institute of Molecular Physics, Russian Research
Center ``Kurchatov Institute'', Kurchatov Sqr. 1, Moscow, 123182
Russia}
\author{O.~Yu.~Gorbenko and A.~R.~Kaul}
\affiliation{Department of Chemistry, Moscow State University,
Vorobievy Gory, Moscow, 119899 Russia}
\author{D.~I.~Khomskii}
\affiliation{Laboratory of Solid State Physics, Materials Science
Center, University of Groningen, Nijenborgh 4, 9747 AG Groningen,
The Netherlands}
\author{K.~I.~Kugel}
\affiliation{Institute for Theoretical and Applied
Electrodynamics, Russian Academy of Sciences, Izhorskaya Str.
13/19, Moscow, 125412 Russia}

\begin{abstract}
The effect of $^{16}$O $\rightarrow$ $^{18}$O isotope substitution
on electrical resistivity and magnetic susceptibility of
Sm$_{1-x}$Sr$_x$MnO$_3$ manganites is analyzed. It is shown that
the oxygen isotope substitution drastically affects the phase
diagram at the crossover region between the ferromagnetic metal
state and that of antiferromagnetic insulator (0.4 $< x <$ 0.6),
and induces the metal-insulator transition at for $x$ = 0.475 and
0.5. The nature of  antiferromagnetic insulator phase is
discussed.
\end{abstract}


\maketitle


Despite intense studies, the nature of the colossal
magnetoresistance in manganites is still a matter of hot debate.
This phenomenon is usually observed when the system is close to a
borderline between ferromagnetic metallic (FM) phase and an
insulating (I) one. The type of the latter may be different
(paramagnetic or antiferromagnetic (AF)), but there are more and
more indications that there exists also some kind of charge
ordering (CO) -- either real long-range CO (see review article
\cite{Rao} and references therein) or, at least, short-range CO
correlations \cite{Lynn,Dai}. When the system is close to a FM-I
crossover, even weak perturbations can induce this crossover:
change of temperature \cite{temper}, magnetic field
\cite{magfield}, pressure \cite{pressure}, irradiation
\cite{irrad}, {\em etc.} In La-Nd-Ca \cite{Zhao} and La-Pr-Ca
\cite{bab1} manganites, one can induce this crossover and
consequently the metal-insulator transition even by changing the
oxygen isotope content: whereas the low-temperature state of
$^{16}$O samples is FM, the samples $^{18}$O become CO AF
insulator. This was confirmed by direct neutron scattering study
\cite{bal}.

The question arises whether this spectacular phenomenon is
confined only to this particular situation of insulating phase
with the charge ordering and the CE-type magnetic structure, i.e.
whether the nature of competing phases, in particular, the
insulating one, is crucial, or one can get similar behavior in
other systems close to the FM-I crossover. In studying this
question, we have found yet another system with the
metal-insulator transition induced by the oxygen isotope
substitution: Sm$_{1-x}$Sr$_x$MnO$_3$ with $x$ in the 0.475--0.5
range. The neutron scattering results for this system at $x$=0.4
\cite{Trounov} and preliminary data for $x$=0.45 with $^{18}$O
\cite{Kurbak} suggest that, in contrast to (La,Pr)CaMnO$_3$, the
insulating phase here has not the CE, but most probably A-type
antiferromagnetic structure. If so, we could conclude that the
metal-insulator transition induced by the isotope substitution is
a general property of a FM-I crossover independent of the detailed
nature of the insulating phase. Let us also note here that X-ray,
neutron, and electron diffraction demonstrated possible existence
of a short-range charge ordering for Sm$_{1-x}$Sr$_x$MnO$_3$ in
the concentration range under discussion \cite{Raveau}. The
results of ESR \cite{Yakub} and Raman \cite{Saitoh} measurements
lead to the same conclusion.


Ceramic Sm$_{1-x}$Sr$_x$MnO$_3$ samples were prepared by the
solid-state reaction technique, the detailed procedure is
described in Ref.~\onlinecite{Aliev}. The enrichment of the
samples by $^{18}$O was performed at $T$ = $950^{\circ }$C and at
pressure  $p$ =1 bar during 200 hours using the method reported in
Refs.~\onlinecite{bab1,bab2}. The resistivity of the samples was
measured by the conventional four-probe technique in the
temperature range from 4.2 to 300 K. The measurements of real part
of ac magnetic susceptibility $\chi'(T)$ were performed in ac
magnetic field with frequency 667~Hz and amplitude of about
0.4~Oe.

\begin{figure}
\includegraphics[height=11 cm,width=7 cm]{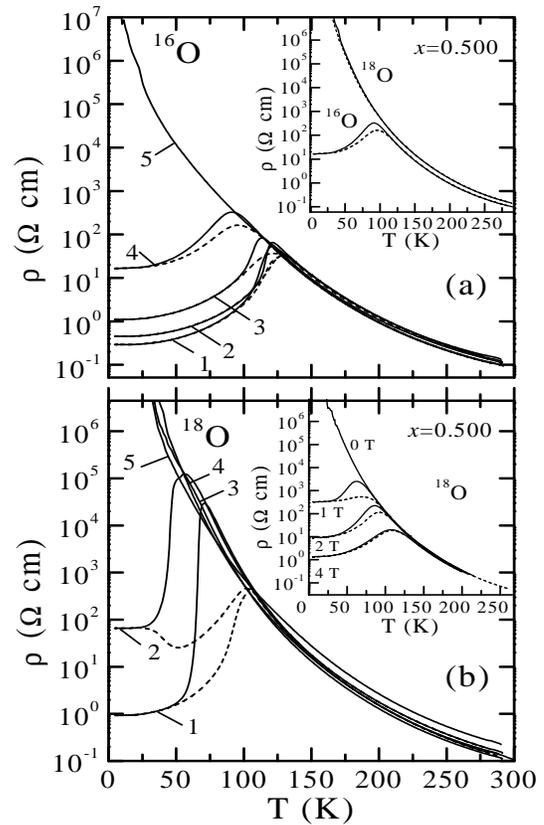}
\caption{\label{Fig_rho} Temperature dependence of electrical
resistivity for  Sm$_{1-x}$Sr$_x$MnO$_3$ with $x$ = 0.425 (1),
0.450 (2), 0.475 (3), 0.500 (4), and 0.525 (5). The results for
samples with $^{16}$O and $^{18}$O are presented at panels (a) and
(b), respectively. The inset in panel (a) illustrates the
metal-insulator transition induced by the oxygen isotope
substitution for samples with $x$ = 0.500. The inset in panel (b)
shows the evolution of resistivity with increasing magnetic field
for the $^{18}$O sample with $x$ = 0.500. Solid and dashed lines
correspond to cooling and heating, respectively.}
\end{figure}

The temperature dependence of electrical resistivity $\rho(T)$ for
Sm$_{1-x}$Sr$_x$MnO$_3$ samples with $x$ = 0.425, 0.450, 0.475,
0.500, and 0.525 annealed both in $^{16}$O and $^{18}$O atmosphere
is presented in Fig.~\ref{Fig_rho}. Four $^{16}$O-containing
samples, with $x$ = 0.425, 0.450, 0.475, and 0.500, are
characterized by a metal-like behavior at low temperatures,
Fig.~\ref{Fig_rho}a. With the growth of $x$, the resistivity
increases and the metal-insulator transition point $T_{MI}$ shifts
toward lower temperatures ($T_{MI}$ was determined as a point
corresponding to the maximum temperature derivative of $\rho(T)$
below the resistivity peak). This can be attributed to the
narrowing of the electron bandwidth and the weakening of
ferromagnetic interaction. The sample with $x$ = 0.525 is an
insulator down to the lowest temperatures.

After the $^{16}$O $\rightarrow$ $^{18}$O isotope substitution,
only two samples with the lowest Sr content ($x$ = 0.425 and
0.450) remain metal-like at low temperatures, the other become
insulating (Fig.~\ref{Fig_rho}b). These metallic samples have much
higher resistivity than those with $^{16}$O. $T_{MI}$ decreases by
40 K and 60 K for the samples with $x$ = 0.425 and 0.450,
respectively. For $x$ = 0.475 and 0.500, we observed the
metal-insulator transition induced by the oxygen isotope
substitution (see inset in Fig.~\ref{Fig_rho}a) similar to that
reported for La-Pr manganites \cite{bab1}. Thus, in this system,
the $^{16}$O $\rightarrow$ $^{18}$O isotope substitution leads to
the changes in the phase diagram: the weakening of the
ferromagnetism and the stabilization of the insulating (probably
CO) phase. A more pronounced thermal hysteresis in $\rho(T)$
curves for the $^{18}$O samples is a manifestation of their
enhanced inhomogeneity. A relatively low magnetic field ($H = 1$
T) transforms the samples with $^{18}$O to the metal-like state
and suppresses the contribution from the high-resistivity
insulating state (for $x$=0.500 see inset in Fig.~\ref{Fig_rho}b).

\begin{figure}
\includegraphics[height=11 cm,width=7 cm]{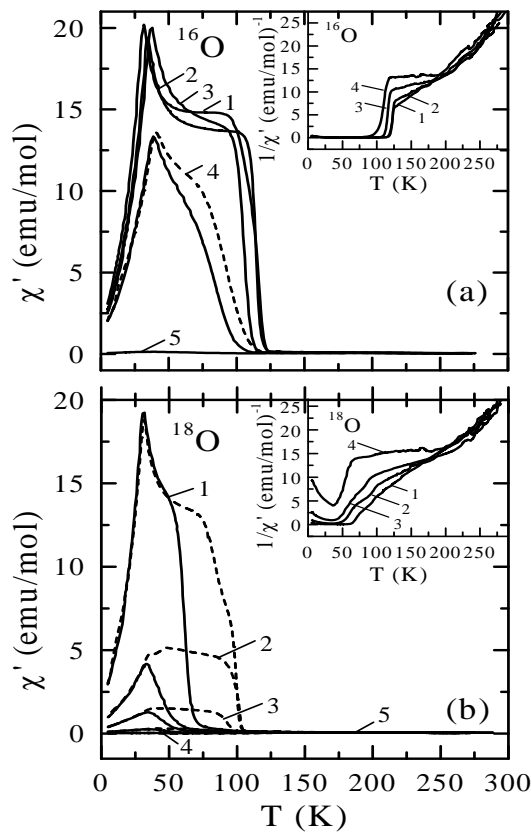}
\caption{\label{Fig_Hi} Temperature dependence of ac magnetic
susceptibility for  Sm$_{1-x}$Sr$_x$MnO$_3$ with $x$ = 0.425 (1),
0.450 (2), 0.475 (3), 0.500 (4), and 0.525 (5). The behavior of
the inverse susceptibility is illustrated in the insets. The
results for samples with $^{16}$O and $^{18}$O are presented at
panels (a) and (b). Solid and dashed lines correspond to cooling
and heating, respectively.}
\end{figure}

The temperature dependence of the ac magnetic susceptibility
$\chi'(T)$ for Sm$_{1-x}$Sr$_x$MnO$_3$ system is presented in
Fig.~\ref{Fig_Hi}. The steep growth of $\chi'(T)$ corresponds to
the onset of FM ordering; the Curie temperature $T_C$ was
determined as a point corresponding to the maximum $d\chi'(T)/dT$.
For samples with $^{16}$O and $x$ = 0.425, 0.450, and 0.475,
$\chi'(T)$ behaves in a similar manner (as far as the magnitude of
$\chi'$ and $T_C$ are concerned). One sees that these samples are
essentially ferromagnetic. The behavior of $\chi'$ at low
temperatures (decrease after reaching a maximum) is apparently
connected with the effect of magnetic domains, see
Ref.~\onlinecite{domains}.

For $x$ = 0.500, the $\chi'$ value drops drastically, $T_C$ shifts
toward lower temperatures, this is a signature of the decreasing
contribution of FM phase. For the composition with $x$ = 0.525 the
susceptibility becomes very small, and the ferromagnetism almost
disappears. In the samples with $^{18}$O, for compositions with $x
\geq$ 0.450 the $\chi'$ value is significantly lower in comparison
to the samples with $^{16}$O, $T_C$ is also much lower and the
hysteresis appears. The FM phase content decreases so steeply that
it becomes insufficient for the percolation in the samples with
$x$ = 0.475 and 0.500, as it is clearly seen for the corresponding
$\rho(T)$ curves. The FM--CO phase boundary shifts toward the CO
state.

The temperature dependence of inverse ac magnetic susceptibility
$1/\chi'(T)$ for Sm$_{1-x}$Sr$_x$MnO$_3$ manganite samples with
$^{16}$O and $^{18}$O is shown in insets in Figs.~\ref{Fig_Hi}a
and \ref{Fig_Hi}b, respectively. Above the magnetic ordering
temperature, the behavior of $1/\chi'(T)$ curves is rather
complicated suggesting the existence of an inhomogeneous state
even in the paramagnetic region. At relatively high temperatures,
$T \sim$ 250--300 K, all $1/\chi'(T)$ are very close to each other
exhibiting a nearly linear growth of the Curie--Weiss type. With
the temperature lowering, the curves flatten, the flattening being
more pronounced for larger $x$. For $x$ = 0.475, $x$ = 0.500, we
can see even a plateau in $1/\chi'(T)$. The onset temperature for
the flattening and deviation of $1/\chi'(T)$ curves for different
$x$ from each other seems to be related to the arising CO
correlations ($T_{CO}\approx 240$ K). This value agrees with the
data of Refs.~\onlinecite{Mukhin,Yakub}. The behavior of
$1/\chi'(T)$ at $T < T_{CO}$ can be interpreted in a picture of
interacting and competing FM and CO correlations, the latter
growing with $x$. Similar behavior was reported in
Ref.~\onlinecite{temper}.

At $T > T_{C}$, the $^{16}$O $\rightarrow ^{18}$O isotope
substitution does not change the general features of $1/\chi'(T)$
curves: the deviations from the Curie--Weiss law begin at the same
temperature, below which the slope of $1/\chi'(T)$ is virtually
independent of the isotope content at a given value of $x$.
However, for the heavier isotope, $T_{C}$ becomes lower and
$1/\chi'$ increases. Thus, for the samples with $^{18}$O, the
plateau in $1/\chi'(T)$ curves is observed within a wider
temperature range. We can argue that the AFM correlations related
to the charge ordering arise in the $^{18}$O samples at the same
temperature $T_{CO} \approx 240$ K, but exist down to lower
temperatures, i.e. the FM interaction becomes weaker and the
CO--FM equilibrium shifts toward the CO state. In the samples with
$^{18}$O, the transition to the FM state is characterized by a
broader hysteresis in comparison to that than for the $^{16}$O
samples.


From the data presented above one sees that the $^{16}$O
$\rightarrow$ $^{18}$O isotope substitution induces the
metal-insulator transition close to the crossover to an insulating
state. According to Ref.~\onlinecite{Raveau}, for $^{16}$O samples
such a crossover starts already at $x$ = 0.4 and extends up to $x$
= 0.6. As follows from our data, the actual change of the ground
state occurs at $x \sim 0.5$ (although an inhomogeneous state with
some traces of the FM phase may exist up to $x$ = 0.6).

The exact nature of an insulating phase for $x \geq$ 0.5 is not
yet established with certainty, but the neutron scattering data
\cite{Kurbak} show that it is an $A$-type antiferromagnet. This
agrees with the general trend characteristic of
R$_{1-x}$Sr$_x$MnO$_3$ manganites where the $A$-type ``bad metal''
state was observed for $x \approx $ 0.5 in compounds with R = La
\cite{LaSr} and R = Nd \cite{KajimotoNdSr}, and with the
theoretical considerations \cite{Ishihara,vdBrinKh}. In our case,
this state is insulating but the resistivity and the energy gap in
it at low temperatures are much smaller than in
(La$_{1-y}$Pr$_y$)$_{0.7}$Ca$_{0.3}$MnO$_3$ with $y$= 0.75, for
which the isotope-induced metal-insulator transition was first
observed in Ref.~\onlinecite{bab1}. Indeed, whereas the room
temperature resistivities of these two systems are comparable
(0.29 Ohm$\cdot$cm in (SmSr)MnO$_3$ {\em vs} 0.37 Ohm$\cdot$cm in
(LaPr)CaMnO$_3$), the resistivities at 60 K are already much
different: 3.5$\cdot$10$^3$ Ohm$\cdot$cm for SmSr system {\em vs}
2.9$\cdot$10$^7$ Ohm$\cdot$cm for (LaPr)Ca one. Note that the
samples of both systems had similar porosity and mean grain size.
The effective activation energies at lowest temperatures can be
estimated as $\sim$15 meV for the SmSr case but $\sim$120 meV for
(LaPr)Ca.

The insulating behavior of SmSr samples be either related to the
granular nature of our samples as contrasted with the single
crystals studied in \cite{Kuwa}, or may be due to a formation of
some weak superstructure of the CO type \cite{Raveau}. As follows
from the results of \cite{Kuwa}, the $A$-type AFM single crystals
show metal-like behavior in the $ab$-plane, but are insulating in
$c$-direction. This can lead to an insulating behavior with small
activation energy in our ceramic samples.

Another factor may be a possible instability of a metallic state.
Most probably, the occupied orbitals in the $A$-type SmSr
insulating phase for $x \geq $ 0.5 are of $x^2-y^2$ type
\cite{KajimotoNdSr,Ishihara, vdBrinKh}, whereas these are ordered
$3x^2-r^2$ and $3y^2-r^2$ in (LaPr)$_{0.7}$Ca$_{0.3}$MnO$_3$, the
latter having the conventional charge ordering with the CE
magnetic structure. The existence of a small energy gap in the
$A$-type SmSr system may be related to the instability of a
metallic state: the Hubbard subband of $x^2-y^2$ type will have
spectrum $\varepsilon (k) = -2t(\cos k_{x}+ \cos k_{y})$ and it
will be half filled at $x$ = 0.5, which would give a nested Fermi
surface. This can lead to a charge-density-wave state with the
opening of a small gap. This should also give a superstructure
with the wavevector  $q = ( \frac{1}{2}, \frac{1}{2}, 0)$ -- the
same as in the usual CO state, but with much weaker distortions.
It would be interesting to look for such weak superstructure
experimentally.

The present results show that the isotope substitution  can
drastically change the properties of the system when it is close
to a FM-I crossover independent of the detailed nature of the I
phase. For the (LaPr)Ca system, this I state is a "strong"
insulator with the charge ordering and the CE magnetic ordering,
whereas here, in SmSr system, it is a weak insulator of the
$A$-type. Nevertheless the effect of the isotope substitution is
similar.

\begin{figure}
\includegraphics[width=8 cm]{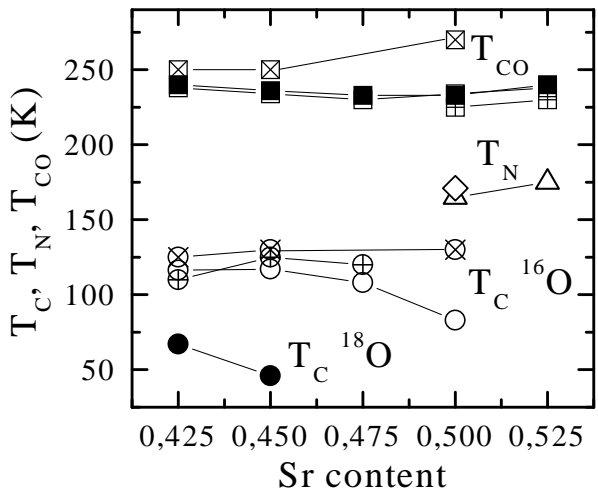}
\caption{\label{fig_Tc} The Curie, N\'{e}el, and charge ordering
temperatures for Sm$_{1-x}$Sr$_x$MnO$_3$ samples under study. Open
and solid symbols correspond to the samples with $^{16}$O and
$^{18}$O, respectively.  $\otimes$, $\lozenge$, $\boxtimes$ and
$\oplus$, $\triangle$, $\boxplus$ are the data from \cite{Yakub}
and \cite{Mukhin} for $T_C$, $T_N$, $T_{CO}$.}
\end{figure}

The detailed mechanism of the isotope effect on the properties of
manganites is not yet completely clear, but most probably it is
connected with the decrease in the electron bandwidth for heavier
isotopes either due to zero-point oscillations or to polaronic
effects \cite{bab2,Edwards}. This is consistent with the change of
$T_{CO}$ and $T_C$ with the isotope content shown in
Fig.~\ref{fig_Tc}. The CO state is relatively insensitive to the
isotope composition, which is quite natural if the mechanism of CO
(or CDW) is predominantly electron-lattice interaction
\cite{KugKhEPL} (as is well known, the dimensionless
electron-phonon coupling constant $\lambda$ does not depend on the
mass of ions). On the contrary, $T_C$ in the double exchange model
scales with the bandwidth, and the decrease in $T_C$ with the
growth of oxygen mass (Fig.~\ref{fig_Tc}) is consistent with this
interpretation.

Summarizing, we studied the effect of the oxygen isotope
substitution on the properties of Sm$_{1-x}$Sr$_x$MnO$_3$ in the
most interesting concentration range 0.4 $\leq x \leq $ 0.6. It
was shown that close to a crossover from the ferromagnetic
metallic state ($x \leq $ 0.5) to an antiferromagnetic insulator
($x \geq $ 0.5) all the properties change drastically -- up to the
fact that for $x$ = 0.475 and 0.5 one even induces the
metal-insulator transition by the $^{16}$O $\rightarrow$ $^{18}$O
substitution. Most probably this substitution transforms the FM
state to an A-type antiferromagnet. We speculate that the relevant
orbitals in this state are $x^2-y^2$ ones and the energy gap
appears due to the formation of a charge density wave in the
$x^2-y^2$ band. We conclude that this transition is induced by the
decrease of the bandwidth for heavier ions. The results obtained,
together with the earlier observed isotope-induced metal-insulator
transition in (La$_{1-y}$Pr$_y$)$_{0.7}$Ca$_{0.3}$MnO$_3$, show
that the isotope substitution is a powerful tool both for
modifying the properties of manganites close to a crossover
between different states and for the study of their physical
characteristics.

We are grateful to A.~N. Taldenkov and A.~V. Inyushkin for helpful
discussions and to A.I. Kurbakov for providing us with his neutron
scattering data prior to publication. The work was supported by
grants of INTAS (01-2008), CRDF (RP2-2355-MO-02), NWO
(097-008-017), and RFBR (01-02-1624, 02-02-16078, 02-03-33258 and
00-15-96570).


\begin{thebibliography}{99}

\bibitem{Rao} C.N.R. Rao {\em et al.},
J. Phys.: Condens. Matter \textbf{12}, R83 (2000).


\bibitem{Lynn} C.P. Adams {\em et al.}, Phys. Rev. Lett., \textbf{85}, 3954 (2000).

\bibitem{Dai} P. Dai {\em et al.}, Phys. Rev. Lett., \textbf{85}, 2553 (2000).

\bibitem{temper} H. Kuwahara {\em et al.}, Phys. Rev. B \textbf{56}, 9386 (1997).

\bibitem{magfield} Y. Tokura {\em et al.}, Phys. Rev. Lett. \textbf{76}, 3184
(1996).

\bibitem{pressure} H. Y. Hwang {\em et al.}, Phys. Rev. B \textbf{52}, 15046 (1995).

\bibitem{irrad} V. Kiryukhin {\em et al.}, Nature (London) \textbf{386}, 813 (1997).

\bibitem{Zhao} Guo-meng Zhao {\em et al.}, Solid State Comm. \textbf{104}, 57 (1997).

\bibitem{bab1} N. A. Babushkina {\em et al.}, Nature (London) \textbf{391}, 159
(1998).

\bibitem{bal} A. M. Balagurov {\em et al.}, Phys. Rev. B. \textbf{60} (1), 383 (1999).

\bibitem{Trounov} I. D. Luzyanin {\em et al.}, Phys. Rev. B \textbf{64}, 094432 (2002).

\bibitem{Kurbak} A. I. Kurbakov, private communication.

\bibitem{Raveau} C. Martin {\em et al.}, Phys. Rev. B. \textbf{60}, 12191 (1999).

\bibitem{Yakub} A. I. Shames {\em et al.}, Solid State Commun. \textbf{121}, 103
(2002).

\bibitem{Saitoh} E. Saitoh {\em et al.}, JMMM \textbf{239}, 170 (2002).

\bibitem{Aliev} A. Aliev {\em et al.}, JETP. Lett. \textbf{72},
464 (2000).

\bibitem{bab2} N. A. Babushkina {\em et al.}, J. Appl.
Phys. \textbf{83}, 7369 (1998).

\bibitem{domains} R. T. Borges {\em et al.}, Phys. Rev. B. \textbf{60}, 12847
(1999).

\bibitem{Mukhin} V. Yu. Ivanov {\em et al.}, JMMM (2002), in press.

\bibitem{LaSr} T. Akimoto {\em et al.}, Phys. Rev. B.
\textbf{57}, R5594 (1998).

\bibitem{KajimotoNdSr} R. Kajimoto {\em et al.}, Phys. Rev. B \textbf{60}
 9506 (1999).

\bibitem{Ishihara} R. Maezono {\em et al.}, Phys. Rev. B.
\textbf{58}, 11583 (1998).

\bibitem{vdBrinKh} J. van den Brink and D. Khomskii, Phys. Rev.
Lett. \textbf{82}, 1016 (1999).

\bibitem{Kuwa} H. Kuwahara {\em et al.}, Phys. Rev. Lett. \textbf{82},4316
(1999).

\bibitem{Edwards} D. M. Edwards, Adv. Phys. (2002), to be published;
cond-mat/0201558.

\bibitem{KugKhEPL}  D.~I.~Khomskii and K.~I.~Kugel, Europhys. Lett.
\textbf{55}, 208 (2001); cond-mat/0112340.



\end{thebibliography}
\end{document}